\documentclass[preprintnumbers,amsmath,amssymbm,prd]{revtex4}
\usepackage{epsfig}
\usepackage{graphicx}
\usepackage{amssymb}

\begin{document}
\title{Natural broadening in the quantum emission spectra of higher-dimensional Schwarzschild black
holes}
\author{Shahar Hod}
\address{The Ruppin Academic Center, Emeq Hefer 40250, Israel}
\address{ }
\address{The Hadassah Institute, Jerusalem 91010, Israel}
\date{\today}

\begin{abstract}
\ \ \ Following an intriguing heuristic argument of Bekenstein, many
researches have suggested during the last four decades that
quantized black holes may be characterized by {\it discrete}
radiation spectra. Bekenstein and Mukhanov (BM) have further argued
that the emission spectra of quantized $(3+1)$-dimensional
Schwarzschild black holes are expected to be {\it sharp} in the
sense that the characteristic natural broadening $\delta\omega$ of
the black-hole radiation lines, as deduced from the quantum
time-energy uncertainty principle, is expected to be much smaller
than the characteristic frequency spacing
$\Delta\omega=O(T_{\text{BH}}/\hbar)$ between adjacent black-hole
quantum emission lines. It is of considerable physical interest to
test the general validity of the interesting conclusion reached by
BM regarding the sharpness of the Schwarzschild black-hole quantum
radiation spectra. To this end, in the present paper we explore the
physical properties of the expected radiation spectra of quantized
$(D+1)$-dimensional Schwarzschild black holes. In particular, we
analyze the functional dependence of the characteristic
dimensionless ratio $\zeta(D)\equiv\delta\omega/\Delta\omega$ on the
number $D+1$ of spacetime dimensions. Interestingly, it is proved
that the dimensionless physical parameter $\zeta(D)$, which
characterizes the sharpness of the black-hole quantum emission
spectra, is an {\it increasing} function of $D$. In particular, we
prove that the quantum emission lines of $(D+1)$-dimensional
Schwarzschild black holes in the regime $D\gtrsim 10$ are
characterized by the dimensionless ratio $\zeta(D)\gtrsim1$ and are
therefore effectively blended together. The results presented in
this paper thus suggest that, even if the underlying energy spectra
of quantized $(D+1)$-dimensional Schwarzschild black holes are
fundamentally {\it discrete}, as argued by many authors, the quantum
phenomenon of natural broadening is expected to smear the
characteristic emission spectra of these higher-dimensional black
holes into a {\it continuum}.
\end{abstract}
\bigskip
\maketitle


\newpage

\section{Introduction}

While studying the interaction of fundamental fields with curved
black-hole spacetimes, Hawking \cite{Haw1} has reached the
remarkable conclusion that, due to quantum effects, black holes are
actually not completely black. In particular, Hawking's seminal
analysis has revealed the intriguing fact that semi-classical black
holes, like mundane black-body emitters, are characterized by {\it
continuous} emission spectra with well defined thermal properties
\cite{Haw1,Notewor}.

Hawking's interesting conclusion regarding the thermal nature of
quantum fields in curved black-hole spacetimes has attracted the
attention of both physicists and mathematicians over the last four
decades and is certainly one of the most important predictions of
fundamental theoretical physics. One should bear in mind, however,
that the ground-breaking analysis presented in \cite{Haw1} has a
fundamentally asymmetric nature: while the fundamental fields living
in the curved black-hole spacetime are properly treated at the {\it
quantum} level, the black hole itself (and, in particular, its
horizon) is treated in \cite{Haw1} as a fixed {\it classical}
entity.

One should therefore regard the continuous black-hole radiation
spectrum derived by Hawking \cite{Haw1} as an important prediction
of {\it semi-classical} general relativity, in which quantized
fundamental fields interact with the classical curved spacetime of a
black hole. Taking cognizance of the fundamental limitations of the
semi-classical theory of general relativity, it is quite natural to
expect that some modifications to the continuous black-hole emission
spectrum predicted by Hawking \cite{Haw1} may arise within the
framework of a self-consistent quantum theory of gravity, a theory
in which the black-hole spacetime itself (and not just the matter
fields) is properly treated as a {\it quantum} physical entity
\cite{Beken1}.

One of the most intriguing quantization schemes for the surface area
(or equivalently, for the energy spectra) of black holes was
presented by Bekenstein more than four decades ago \cite{Beken1}.
Following the interesting physical observation that the surface area
of a non-extremal black hole behaves as a fundamental adiabatic
invariant quantity \cite{Beken1,Cris}, Bekenstein has argued, using
the Ehrenfest principle \cite{Ehr}, that the surface area of a
quantized black hole should be characterized by a uniformly spaced
{\it discrete} spectrum of the form \cite{Beken1,Noteunit}
\begin{equation}\label{Eq1}
A_n=4\gamma\hbar\cdot n\ \ \ ;\ \ \ n\in\mathbb{Z}\  ,
\end{equation}
where $\gamma$ is a dimensionless constant of order unity. The three
most commonly used values of the parameter $\gamma$ which appear in
the physics literature are: $\gamma=2\pi$ \cite{Beken1},
$\gamma=\ln2$ \cite{BekMuk,Hodch}, and $\gamma=\ln3$ \cite{Hodln}.
Interestingly, the remarkably compact formula (\ref{Eq1}) suggested
by Bekenstein \cite{Beken1} for the discrete surface area of
$(3+1)$-dimensional quantized black holes has been re-derived by
several authors who have used different physically motivated
quantization schemes (see
\cite{BekMuk,Hodch,Hodln,Muk,Kog,Maz,Mag,Lou,Pel,Louk,Bar,Kas,MakRep,HodD,HodG,Boj,Alu,Gar1,Gar2,Magim,Kund,Kes}
and references therein).

Bekenstein has further argued \cite{Beken1} that the uniformly
spaced area spectrum (\ref{Eq1}) should be associated with a
discrete mass (energy) spectrum $\{M_n\}$ for quantized black holes.
In particular, using the simple mass-area relation
\begin{equation}\label{Eq2}
A=16\pi M^2
\end{equation}
for $(3+1)$-dimensional Schwarzschild black holes, one immediately
deduces from (\ref{Eq1}) a discrete mass spectrum of the form
\begin{equation}\label{Eq3}
M_n=\sqrt{{{\gamma\hbar}\over{4\pi}}\cdot n}\ \ \ ;\ \ \
n\in\mathbb{Z}\
\end{equation}
for these spherically symmetric black holes. Taking cognizance of
the discrete energy spectrum (\ref{Eq3}), Bekenstein and Mukhanov
(BM) \cite{Beken1,BekMuk} have raised the intriguing idea that,
within the framework of a quantum theory of gravity, quantized
Schwarzschild black holes may be characterized by {\it discrete}
radiation spectra.

In particular, as stressed by BM \cite{Beken1,BekMuk}, the decay of
a macroscopic \cite{Notemac} $(3+1)$-dimensional quantized
Schwarzschild black hole of mass $M_n$ into lower energy levels is
expected to be accompanied by the emission of discrete field quanta
whose characteristic frequencies are given by \cite{Notenk}
\begin{equation}\label{Eq4}
\omega_k={{M_n-M_{n-k}}\over{\hbar}}=k\cdot\varpi\ \ \ \ ; \ \ \ \
k=1,2,3,...\ ,
\end{equation}
where the fundamental ({\it smallest} possible) frequency $\varpi$
which characterizes the quantized black-hole emission spectrum is
given by the simple dimensionless relation [see Eq. (\ref{Eq3})]
\begin{equation}\label{Eq5}
M\varpi={{\gamma}\over{8\pi}}\  .
\end{equation}
Interestingly, the quantized radiation spectrum (\ref{Eq4})
advocated by BM \cite{Beken1,BekMuk} for macroscopic \cite{Notemac}
$(3+1)$-dimensional Schwarzschild black holes is characterized by
the remarkably simple constant spacing \cite{Notespa}
\begin{equation}\label{Eq6}
\Delta\omega=\varpi\
\end{equation}
between the corresponding frequencies of adjacent black-hole
emission lines.

\section{Natural broadening of the quantized black-hole emission lines}

In order to establish the discrete nature of the proposed radiation
spectrum (\ref{Eq4}) of a quantized $(3+1)$-dimensional
Schwarzschild black hole, Bekenstein and Mukhanov \cite{BekMuk} have
analyzed the influence of the quantum phenomenon of {\it natural
broadening} \cite{Ehr} on the widths of the black-hole emission
lines. In particular, using the time-energy quantum uncertainty
principle \cite{Ehr}, BM \cite{BekMuk} (see also \cite{Mak,Hod1})
have related the natural frequency broadening $\delta\omega$ of the
black-hole quantum emission lines to the reciprocal of the
characteristic {\it finite} lifetime $\tau$ of the black-hole $n$th
energy (mass) level \cite{Ehr,Notetua}:
\begin{equation}\label{Eq7}
\delta\omega={{1}\over{\tau}}\  .
\end{equation}
In the spirit of the Bohr correspondence principle \cite{Ehr},
Bekenstein and Mukhanov \cite{BekMuk} have further suggested to
relate the characteristic lifetime $\tau$ of the $n$th energy (mass)
level of a macroscopic \cite{Notemac} quantized $(3+1)$-dimensional
Schwarzschild black hole to the reciprocal of the corresponding
semi-classical emission rate which characterizes the black hole
\cite{Haw1,Page,Notecor,Jack,Hor}:
\begin{equation}\label{Eq8}
\tau=\Big({{dN}\over{dt}}\Big)^{-1}\  .
\end{equation}

The sharpness of the black-hole emission spectrum (\ref{Eq4}) can be
characterized by the dimensionless ratio $\delta\omega/\Delta\omega$
between the natural frequency width of the spectral lines [see Eq.
(\ref{Eq7})] and the characteristic frequency spacing between
adjacent radiation lines [see Eqs. (\ref{Eq5}) and (\ref{Eq6})]. In
particular, {\it discrete} emission spectra are characterized by the
strong inequality $\delta\omega/\Delta\omega\ll1$, whereas emission
spectra which are effectively {\it continuous} are characterized by
the strong inequality $\delta\omega/\Delta\omega\gg1$.

Interestingly, one finds \cite{BekMuk,Hod1} that the emission
spectrum of a quantized $(3+1)$-dimensional Schwarzschild black hole
is characterized by the relation [see Eq. (\ref{Eq20}) below]
\begin{equation}\label{Eq9}
\zeta(D=3)\equiv{{{\delta\omega}}\over{{\Delta\omega}}}\ll1\ .
\end{equation}
As emphasized by BM \cite{BekMuk}, the small ratio (\ref{Eq9}) found
for the dimensionless physical parameter $\zeta(D=3)$ implies that
the discrete eigenfrequencies (\ref{Eq4}), which according to
Bekenstein \cite{Beken1} (see also
\cite{BekMuk,Hodch,Hodln,Muk,Kog,Maz,Mag,Lou,Pel,Louk,Bar,Kas,MakRep,HodD,HodG,Boj,Alu,Gar1,Gar2,Magim,Kund,Kes})
are expected to characterize the radiation spectra of quantized
$(3+1)$-dimensional Schwarzschild black holes, are unlikely to
overlap.

\section{The spectral emission lines of $(D+1)$-dimensional Schwarzschild black holes
and their natural quantum broadening}

It is of physical interest to test the general validity of the
intriguing conclusion reached by BM \cite{BekMuk}, according to
which the quantum phenomenon of natural broadening has a {\it
negligible} effect on the suggested discrete emission spectra
(\ref{Eq4}) of quantized $(3+1)$-dimensional Schwarzschild black
holes. In particular, one naturally wonders whether the strong
inequality $\delta\omega\ll\Delta\omega$ [see Eq. (\ref{Eq9})],
which characterizes the emission spectra of quantized
$(3+1)$-dimensional Schwarzschild black holes, is a generic property
of {\it all} $(D+1)$-dimensional quantized Schwarzschild black
holes?

In order to address this physically interesting question, in the
present paper we shall analyze the functional dependence of the
dimensionless physical ratio
\begin{equation}\label{Eq10}
\zeta(D)\equiv{{\delta\omega(D)}\over{\Delta\omega(D)}}\  ,
\end{equation}
which quantifies the sharpness \cite{Noteshar} of the black-hole
quantum emission spectra, on the number $D+1$ of spacetime
dimensions.

It is worth emphasizing that, as extensively discussed in the
literature (see \cite{Ark,Ran,Cas,Ear} and references therein),
higher-dimensional physical theories with extra spatial dimensions
provide intriguing candidates for self-consistent physical theories
unifying the fundamental forces of nature. Interestingly, the
suggested higher-dimensional physical theories
\cite{Ark,Ran,Cas,Ear} may provide an elegant resolution for the
hierarchy problem observed in our universe. Moreover, physical
theories with extra spatial dimensions predict the formation of mini
black holes in future high-energy accelerators
\cite{Ark,Ran,Cas,Ear}. Interestingly, these higher-dimensional mini
black holes are expected to be characterized by quantum emission
spectra. It is therefore of physical interest to explore the
physical properties (and, in particular, the characteristic quantum
emission spectra) of these predicted higher-dimensional black holes
that will hopefully be observed in future man-made high-energy
scattering experiments.

\subsection{The emission spectra of quantized $(D+1)$-dimensional
Schwarzschild black holes}

Interestingly, the quantization schemes presented in
\cite{Beken1,Hodln,Kunl} suggest that higher-dimensional
Schwarzschild black holes, like their $(3+1)$-dimensional
counterparts, are expected to be characterized by evenly spaced
discrete emission spectra of the form
\begin{equation}\label{Eq11}
\omega_k=k\cdot\varpi(D)\ \ \ \ ; \ \ \ \ k=1,2,3,...\ ,
\end{equation}
where the fundamental radiation frequency of a quantized
$(D+1)$-dimensional Schwarzschild black hole is given by the compact
physical expression \cite{Kunl,Notehgs}
\begin{equation}\label{Eq12}
\varpi(D)={{\gamma T_{\text{BH}}(D)}\over{\hbar}}\  .
\end{equation}
Here \cite{SchTang,Noterh}
\begin{equation}\label{Eq13}
T_{\text{BH}}(D)={{(D-2)\hbar}\over{4\pi r_{\text{H}}}}\
\end{equation}
is the characteristic Bekenstein-Hawking temperature of a
$(D+1)$-dimensional Schwarzschild black hole of horizon radius
$r_{\text{H}}$.

Note that the quantized radiation spectrum (\ref{Eq11}), suggested
for macroscopic \cite{Notemac} $(D+1)$-dimensional Schwarzschild
black holes by the quantization schemes of \cite{Beken1,Hodln,Kunl},
is characterized by the constant frequency spacing \cite{Notespa}
\begin{equation}\label{Eq14}
\Delta\omega(D)=\varpi(D)\
\end{equation}
between adjacent black-hole spectral lines.

\subsection{Natural broadening and radiation fluxes of $(D+1)$-dimensional
Schwarzschild black holes}

Following the physical procedure suggested by Bekenstein and
Mukhanov \cite{BekMuk} (see also \cite{Mak,Hod1}), we shall
determine the natural frequency broadening $\delta\omega(D)$ of the
spectral lines (\ref{Eq11}), which are expected to characterize the
emission spectra of macroscopic $(D+1)$-dimensional quantized
Schwarzschild black holes, from the characteristic relation [see
Eqs. (\ref{Eq7}) and (\ref{Eq8})]
\begin{equation}\label{Eq15}
\delta\omega(D)={{dN}\over{dt}}\  ,
\end{equation}
where ${{dN}/{dt}}$ is the corresponding $(D+1)$-dimensional
semi-classical emission rate (that is, the number of quanta emitted
per unit of time) of the higher-dimensional Schwarzschild black hole
\cite{Haw1,Page,Notecor,Jack,Hor,Notehb}.

Below we shall consider the emission of massless gravitons and
photons \cite{Haw1,Page,ZuKa,CKK}. In particular, in the present
analysis we shall assume that the radiating $(D+1)$-dimensional
Schwarzschild black holes are macroscopic in the sense that the
emission of massive particles in the regime $\mu\cdot
r_{\text{H}}\gg D^2\hbar$ is exponentially suppressed
\cite{Page,ZuKa}. [Note, in particular, that an emitted massive
particle can reach spatial infinity only if its proper energy
satisfies the inequality $\hbar\omega\geq\mu$. The exponential
factor $\omega^{D-1}/(e^{\hbar\omega/T_{\text{BH}}}-1)$ that governs
the radiation flux of a $(D+1)$-dimensional Schwarzschild black hole
[see Eq. (\ref{Eq16}) below] implies that the corresponding
radiation rate of massive quanta in the regime
$\hbar\omega\geq\mu\gg D^2\hbar/r_{\text{H}}$ is exponentially
suppressed]. Furthermore, it should be noted that, had we considered
an extended set of emitted particles (which includes massive
particles along with massless fields), we would have found {\it
shorter} lifetimes for the meta-stable (radiating) black-hole
states. Thus, extending the family of emitted field modes would
merely strengthen our final conclusion [see Eqs. (\ref{Eq21}) and
(\ref{Eq23}) below] that, in the large-D regime, the quantum
phenomenon of natural broadening \cite{Ehr} is expected to smear the
characteristic emission spectra of radiating $(D+1)$-dimensional
Schwarzschild black holes into a continuum.

The semi-classical radiation flux out of a $(D+1)$-dimensional
Schwarzschild black hole for one massless bosonic degree of freedom
is given by the expression \cite{Haw1,Page,ZuKa,CKK}
\begin{equation}\label{Eq16}
{{dN}\over{dt}}={{1}\over{2^{D-1}\pi^{D/2}\Gamma(D/2)}}\sum_{j}{\int_0^{\infty}}\Gamma
{{\omega^{D-1}}\over{{e^{\hbar\omega/T_{\text{BH}}}-1}}} d\omega\  ,
\end{equation}
where $j$ denotes the angular harmonic parameters of the emitted
field quanta. The absorption probabilities
$\Gamma=\Gamma(\omega;j,D)$ (known as the black-hole-field greybody
factors) \cite{Page} that appear in the integral relation
(\ref{Eq16}) quantify the linearized interaction of the emitted
field modes with the effective gravitational potential of the curved
$(D+1)$-dimensional Schwarzschild black-hole spacetime. These
characteristic black-hole-field reflection coefficients are
determined by solving a standard problem of wave scattering in the
curved black-hole spacetime. In particular, the linearized
interaction (scattering) of the emitted field modes with the curved
black-hole spacetime is governed by the generalized
$(D+1)$-dimensional Regge-Wheeler equation \cite{Haw1,Page,ZuKa,CKK}
\begin{equation}\label{Eq17}
\Big({{d^2}\over{dx^2}}+\omega^2-V\Big)\phi=0\  ,
\end{equation}
where the radial coordinate $x$ is related to the Schwarzschild coordinate $r$ by the differential
relation $dx/dr=[1-(r_{\text{H}}/r)^{D-3}]^{-1}$. For a massless
perturbation field of harmonic index $l$, the effective
$(D+1)$-dimensional black-hole curvature potential in (\ref{Eq17})
is given by the cumbersome expression \cite{CKK}
\begin{equation}\label{Eq18}
V(r;D)=\Big[1-\Big({{r_{\text{H}}}\over{r}}\Big)^{D-3}\Big]\Big[{{l(l+D-2)+(D-1)(D-3)/4}\over{r^2}}
+{{(1-p^2)(D-1)^2r^{D-2}_{\text{H}}}\over{4r^D}}\Big]\  .
\end{equation}
Here one should take the values $p=\{0,2,2/(D-1),2(D-2)/(D-1)\}$ for
the distinct cases of gravitational tensor fields, gravitational
vector fields, electromagnetic vector fields, and electromagnetic
scalar fields, respectively \cite{CKK} [It should be noted that the
effective radial potential in the generalized $(D+1)$-dimensional
Regge-Wheeler equation (\ref{Eq17}) for the case of gravitational
scalar perturbation fields is characterized by a rather complicated
expression which is given in \cite{CKK}].

\section{Sharpness of the $(D+1)$-dimensional
Schwarzschild black-hole emission spectra: Numerical and analytical
results}

In the present section we shall explore the functional dependence of
the dimensionless physical parameter
$\zeta(D)\equiv{{\delta\omega(D)}/{\Delta\omega(D)}}$ [see Eqs.
(\ref{Eq14}) and (\ref{Eq15})], which characterizes the sharpness
\cite{Noteshar} of the black-hole quantum emission spectra, on the
number $D+1$ of spacetime dimensions.

\subsection{The $(3+1)$-dimensional Schwarzschild black hole}

The total emission rate of massless gravitons and photons from a
macroscopic $(3+1)$-dimensional Schwarzschild black hole is given by
\cite{Page} $dN/dt\simeq 1.6\times 10^{-4}M^{-1}$. Using the
characteristic relation (\ref{Eq15}), one finds
\begin{equation}\label{Eq19}
\delta\omega(D=3)\simeq1.6\times 10^{-4}M^{-1}
\end{equation}
for the natural frequency broadening which characterizes the
emission lines of the quantized $(3+1)$-dimensional Schwarzschild
black holes.

Taking cognizance of Eqs. (\ref{Eq5}), (\ref{Eq6}), and
(\ref{Eq19}), one finds that the emission spectra of quantized
$(3+1)$-dimensional Schwarzschild black holes are characterized by
the remarkably small dimensionless ratio \cite{Notegam,NoteeKr}
\begin{equation}\label{Eq20}
\zeta(D=3)\equiv{{\delta\omega(D=3)}\over{\Delta\omega(D=3)}}\simeq
4\times 10^{-3}\ll 1\  .
\end{equation}
The extremely small value (\ref{Eq20}) found for the dimensionless
physical parameter $\zeta(D=3)$ implies that the quantum phenomenon
of natural broadening has a negligible effect on the expected
emission spectra of quantized $(3+1)$-dimensional Schwarzschild
black holes. In particular, as emphasized by BM \cite{BekMuk}, the
characteristic strong inequality
$\delta\omega(D=3)\ll\Delta\omega(D=3)$ implies that the discrete
emission frequencies (\ref{Eq4}), which according to Bekenstein
\cite{Beken1} (see also
\cite{BekMuk,Hodch,Hodln,Muk,Kog,Maz,Mag,Lou,Pel,Louk,Bar,Kas,MakRep,HodD,HodG,Boj,Alu,Gar1,Gar2,Magim,Kund,Kes})
are expected to characterize the radiation spectra of quantized
$(3+1)$-dimensional Schwarzschild black holes, are unlikely to
overlap.

\subsection{$(D+1)$-dimensional Schwarzschild black holes: Intermediate
D-values}

In the previous subsection we have seen that $(3+1)$-dimensional
quantized Schwarzschild black holes are characterized by a
remarkably small value of the physical parameter $\zeta(D=3)$. In
the present subsection we shall explicitly prove that the
fundamental dimensionless ratio $\zeta(D)$, which characterizes the
sharpness \cite{Noteshar} of the black-hole quantum emission
spectra, is an increasing function of the spacetime dimension $D+1$.

The emission rates of massless gravitons and photons from
$(D+1)$-dimensional Schwarzschild black holes were computed
numerically in \cite{CKK}. In Table \ref{Table1} we display, for
intermediate values of the black-hole spacetime dimension $D+1$, the
numerically computed values of the dimensionless physical parameter
$\zeta(D)\equiv\delta\omega/\Delta\omega$ [see Eqs.
(\ref{Eq12})-(\ref{Eq15})] \cite{Notegam}. The data presented in
Table \ref{Table1} reveal the intriguing fact that the physical
parameter $\zeta(D)$, which quantifies the sharpness of the
quantized $(D+1)$-dimensional Schwarzschild black-hole emission
spectra, is an {\it increasing} function of the number $D+1$ of
spacetime dimensions.

In particular, one finds from Table \ref{Table1} that
higher-dimensional Schwarzschild black holes in the regime $D\gtrsim
10$ are characterized by the inequality \cite{Noteex}
\begin{equation}\label{Eq21}
\delta\omega\gtrsim\Delta\omega\ \ \ \ \text{for}\ \ \ \ D\gtrsim10\
.
\end{equation}
The relation (\ref{Eq21}) strongly suggests that, due to the quantum
phenomenon of natural broadening \cite{Ehr}, the characteristic
spectral lines (\ref{Eq11}) of quantized higher-dimensional
Schwarzschild black holes are expected to be effectively blended
together in the regime $D\gtrsim 10$.

\begin{table}[htbp]
\centering
\begin{tabular}{|c|c|c|c|c|c|c|c|}
\hline $D+1$ & \ \ 5 \ \ & \ \ 6 \ \ & \ \ 7 \ \ & \ \ 8 \ \ & \ \ 9 \ \ & \ \ 10 \ \ & \ \ 11\ \ \ \\
\hline \ \ $\zeta(D)\equiv\delta\omega/\Delta\omega$ \ \ &\ \ \
0.046\ \ \ \ &\ \ \ 0.151\ \ \ \ &\ \ \ 0.310\ \ \ \ &\ \ \ 0.645\ \
\ \ &\ \ \ 1.036\ \ \ \ &\ \ \ 2.153\ \ \ \ &\ \ \ 3.639\ \ \ \ \\
\hline
\end{tabular}
\caption{The dimensionless physical parameter
$\zeta(D)\equiv\delta\omega/\Delta\omega$ which quantifies the
sharpness of the $(D+1)$-dimensional Schwarzschild black-hole
quantum emission spectra. Here $\delta\omega$ is the natural
broadening of the black-hole quantum emission lines [see Eq.
(\ref{Eq15})] and $\Delta\omega$ \cite{Notegam} is the
characteristic frequency spacing between adjacent emission lines of
the quantized $(D+1)$-dimensional Schwarzschild black holes [see
Eqs. (\ref{Eq12})-(\ref{Eq14})]. One finds that the dimensionless
physical parameter $\zeta(D)$ is an {\it increasing} function of the
number $D+1$ of spacetime dimensions. In particular, we find that
higher-dimensional Schwarzschild black holes in the regime $D=O(10)$
are characterized by the relation $\delta\omega/\Delta\omega=O(1)$
\cite{Noteex}.} \label{Table1}
\end{table}

\subsection{$(D+1)$-dimensional Schwarzschild black holes: The large-$D$ regime}

In the previous subsection we have used numerical data to reveal the
interesting fact that the dimensionless physical parameter
$\zeta=\zeta(D)$, which quantifies the sharpness of the black-hole
quantum emission lines, is an increasing function of the number
$D+1$ of spacetime dimensions. In the present subsection we shall
use analytical results in order to prove that this fundamental
physical parameter is characterized by the asymptotic behavior
$\zeta(D\gg1)\gg1$.

It has recently been demonstrated explicitly \cite{Hoddd} that the
semi-classical radiation spectra \cite{Haw1,Page,Notecor,Jack,Hor}
of $(D+1)$-dimensional Schwarzschild black holes in the asymptotic
large-D regime are described remarkably well by the eikonal
(short-wavelengths) approximation. In particular, using the
geometric-optics approximation, one finds \cite{Hoddd} the
remarkably compact analytical formula
\begin{equation}\label{Eq22}
{{dN}\over{dt}}\times
r_{\text{H}}={{(4\pi)^2}\over{e}}\Big({{D}\over{4\pi}}\Big)^{D+3}\ \
\ \text{for}\ \ \ D\gg1\
\end{equation}
for the dimensionless semi-classical radiation flux of a
$(D+1)$-dimensional Schwarzschild black hole in the asymptotic
large-D regime.

Taking cognizance of Eqs. (\ref{Eq12}), (\ref{Eq13}), (\ref{Eq14}),
(\ref{Eq15}), and (\ref{Eq22}), one deduces that the emission
spectra of quantized higher-dimensional Schwarzschild black holes in
the large-D regime are expected to be characterized by the
dimensionless asymptotic relation \cite{Noteasy,Notemat}
\begin{equation}\label{Eq23}
\zeta(D)={{(4\pi)^2}\over{\gamma
e}}\Big({{D}\over{4\pi}}\Big)^{D+2}\ \ \ \text{for}\ \ \ D\gg1\  .
\end{equation}
The characteristic strong inequality $\delta\omega\gg\Delta\omega$
found in the large-D regime [see Eqs. (\ref{Eq10}) and (\ref{Eq23})]
strongly suggests that the quantum phenomenon of natural broadening
\cite{Ehr} would effectively smear the corresponding radiation lines
of these higher-dimensional Schwarzschild black holes into a
continuum.

\section{Summary and discussion}

Following the highly influential work of Bekenstein \cite{Beken1},
many researches (see
\cite{BekMuk,Hodch,Hodln,Muk,Kog,Maz,Mag,Lou,Pel,Louk,Bar,Kas,MakRep,HodD,HodG,Boj,Alu,Gar1,Gar2,Magim,Kund,Kes}
and references therein) have argued during the last four decades
that, within the framework of a self-consistent quantum theory of
gravity \cite{Noteqg}, black holes should be characterized by {\it
discrete} radiation spectra with evenly spaced spectral lines [see
Eq. (\ref{Eq4})].

Furthermore, using the quantum time-energy uncertainty principle
\cite{Ehr}, Bekenstein and Mukhanov \cite{BekMuk} have reached the
important physical conclusion that the characteristic radiation
spectra of quantized $(3+1)$-dimensional Schwarzschild black holes
are expected to be {\it sharp} in the sense that the characteristic
natural broadening $\delta\omega$ [see Eqs. (\ref{Eq7}) and
(\ref{Eq8})] of the black-hole quantum emission lines is much
smaller than the characteristic spacing
$\Delta\omega=O(T_{\text{BH}}/\hbar)$ [see Eqs. (\ref{Eq5}) and
(\ref{Eq6})] between adjacent emission lines of the quantized black
holes. It was therefore concluded by BM \cite{BekMuk} that, for
quantized $(3+1)$-dimensional Schwarzschild black holes, the
characteristic discrete spectral lines (\ref{Eq4}), as predicted in
\cite{Beken1,BekMuk,Hodch,Hodln,Muk,Kog,Maz,Mag,Lou,Pel,Louk,Bar,Kas,MakRep,HodD,HodG,Boj,Alu,Gar1,Gar2,Magim,Kund,Kes},
are unlikely to overlap.

One naturally wonders whether the strong inequality
$\delta\omega(D=3)\ll\Delta\omega(D=3)$ \cite{BekMuk}, which
characterizes the expected spectral lines (\ref{Eq4}) of quantized
$(3+1)$-dimensional Schwarzschild black holes, is a generic property
of {\it all} quantized $(D+1)$-dimensional Schwarzschild black
holes? In order to address this physically interesting question, in
the present paper we have studied the characteristic radiation
spectra of $(D+1)$-dimensional quantized Schwarzschild black holes.
In particular, we have analyzed the functional dependence of the
characteristic dimensionless ratio
$\zeta(D)\equiv\delta\omega/\Delta\omega$ on the spacetime dimension
$D+1$ of the quantized black hole. Interestingly, we have explicitly
proved that the dimensionless physical parameter $\zeta(D)$, which
quantifies the natural broadening (the sharpness) of the black-hole
quantum emission lines, is an {\it increasing} function of the
number $D+1$ of spacetime dimensions (see Table \ref{Table1}).

In particular, we have shown that the quantum emission lines of
$(D+1)$-dimensional Schwarzschild black holes in the regime
$D\gtrsim10$ are characterized by the dimensionless ratio
$\zeta(D)\gtrsim1$ [see Eq. (\ref{Eq21})]. Moreover,
we have proved that the emission spectra of quantized
$(D+1)$-dimensional Schwarzschild black holes are characterized by
the large-D asymptotic behavior $\zeta(D\to\infty)\to\infty$ [see
Eq. (\ref{Eq23})] \cite{Noteasy}. These intriguing findings imply,
in particular, that the characteristic emission lines of these
higher-dimensional quantized black holes are effectively blended
together.

The results presented in the present paper therefore suggest that,
even if the underlying energy spectra of quantized
$(D+1)$-dimensional Schwarzschild black holes are fundamentally {\it
discrete}, as argued by many authors
\cite{Beken1,BekMuk,Hodch,Hodln,Muk,Kog,Maz,Mag,Lou,Pel,Louk,Bar,Kas,MakRep,HodD,HodG,Boj,Alu,Gar1,Gar2,Magim,Kund,Kes},
the quantum phenomenon of natural broadening \cite{Ehr} is expected
to smear the characteristic emission spectra of these
higher-dimensional black holes into a {\it continuum}.

\bigskip
\noindent
{\bf ACKNOWLEDGMENTS}
\bigskip

This research is supported by the Carmel Science Foundation. I thank
Yael Oren, Arbel M. Ongo, Ayelet B. Lata, and Alona B. Tea for
stimulating discussions.



\begin{thebibliography}{99}
\bibitem{Notewor} It is worth emphasizing the fact that the characteristic thermal
emission spectra of semi-classical black holes are partially
distorted by the effective curvature (gravitational) potentials [see
Eq. (\ref{Eq18}) below] which quantify the linearized interaction of
the emitted field modes with the corresponding curved black-hole
spacetimes.
\bibitem{Noteunit} We shall use natural units in which $G=c=k_{\text{B}}=1$.
\bibitem{Notemac} Here we use the term `macroscopic black hole' to describe a black hole
whose surface area is characterized by the strong inequality
$A_n\gg\hbar$ [that is, $n\gg1$, see Eq. (\ref{Eq1})].
\bibitem{Notenk} Here one assumes that the integer $k$, which characterizes the energy
transition of the quantized black hole, satisfies the relations
$1\leq k\ll n$ [note that $n\gg1$ for macroscopic black holes, see
Eq. (\ref{Eq1}) and \cite{Notemac}].
\bibitem{Notespa} The characteristic frequency spacing between adjacent emission lines
is given by $\Delta\omega\equiv \omega_{k+1}-\omega_k$ [see Eq.
(\ref{Eq4})].
\bibitem{Notetua} That is, $\tau$ is the average time (as measured by asymptotic observers)
between physical transitions (quantum leaps) of the black hole from
a given energy (mass) level to a lower energy level.
\bibitem{Notehgs} Note that the expression (\ref{Eq12}) for the fundamental emission
frequency of a quantized $(D+1)$-dimensional Schwarzschild black
hole reduces to (\ref{Eq5}) in the familiar $(3+1)$-dimensional
case.
\bibitem{Noterh} Note that a spherically symmetric $(D+1)$-dimensional Schwarzschild black hole
is characterized by an horizon radius of the form
$r_{\text{H}}={[{{16\pi M}/{(D-1)A_{D-1}}}]}^{1/{(D-2)}}$
\cite{SchTang}, where $M$ is the black-hole mass and
$A_{D-1}={{2\pi^{D/2}}/{\Gamma(D/2)}}$ is the generalized surface
area of a unit $(D-1)$-dimensional sphere.
\bibitem{Notehb} It should be noted that, in order to estimate the characteristic emission rate of
a $(3+1)$-dimensional Schwarzschild black hole, BM \cite{BekMuk}
have used an approximated approach in which the radiating black hole
was modeled as a perfect black-body emitter. In the present paper we
shall consider the exact radiation fluxes of $(D+1)$-dimensional
Schwarzschild black holes, taking into account the partial
reflection of the emitted field modes by the gravitational potential
[see Eq. (\ref{Eq18}) below] of the curved black-hole spacetime.
\bibitem{Notegam} Here we have used the value $\gamma=1$.
This order of magnitude for
the dimensionless physical parameter $\gamma$ is consistent with the
values most commonly used in the physical literature for this
fundamental coefficient, see \cite{Beken1,BekMuk,Hodch,Hodln}.
\bibitem{NoteeKr} See \cite{Hod1} for the analysis of the natural broadening phenomenon
which is expected to characterize the emission spectra of quantized
rapidly-rotating $(3+1)$-dimensional Kerr black holes.
\bibitem{Noteex} Note that the exact number $D+1$ of spacetime dimensions for which
$\zeta(D)=1$ depends on the chosen value
\cite{Beken1,BekMuk,Hodch,Hodln} of the dimensionless physical
parameter $\gamma$.
\bibitem{Noteasy} It is worth emphasizing the fact that, from the analytically
derived expression (\ref{Eq23}), one finds the asymptotic behavior
$\zeta(D\to\infty)\to\infty$ for the fundamental physical parameter
$\zeta$. Note, in particular, that this characteristic asymptotic
behavior of the dimensionless ratio $\delta\omega/\Delta\omega$ is
{\it independent} of the value of the dimensionless constant
$\gamma$.
\bibitem{Notemat} It is worth noting that the analytically derived asymptotic formula (\ref{Eq23}),
which is based on the geometric-optics (eikonal) approximation in
the asymptotic large-D regime, provides the correct order of
magnitude for the dimensionless physical parameter $\zeta(D)$
already for $D=10$. Specifically, one finds from (\ref{Eq23})
$\zeta^{\text{analytical}}(D=10)=3.746$ \cite{Notegam}, which is a
factor $\sim 1.03$ larger than the corresponding numerically
computed value $\zeta^{\text{numerical}}(D=10)\simeq 3.639$ (see
Table \ref{Table1}).
\end{thebibliography}
\end{document}